\documentclass[a4paper,conference]{IEEEtran}
\usepackage{research5IEEE}
\usepackage{todonotes}

\usepackage{mathrsfs,url,psfrag,graphicx,epsfig,mathtools,research5IEEE,amsmath,upgreek,
	amsfonts,amssymb,amsthm,tikz,array,wasysym,pifont,dsfont,balance}

\usetikzlibrary{arrows}

\newcommand{\GameNF}{\mathcal{G}(b) = \left(\mathcal{K}, \left\lbrace\mathcal{A}_k \right\rbrace_{k \in \mathcal{K}},\left\lbrace
	u_{k}\right\rbrace_{ k \in \mathcal{K}}\right)}

\newcommand{\outage}{\textnormal{outage}}
\IEEEoverridecommandlockouts

\newcommand{\gameNF}{\mathcal{G}(b)}

\newtheorem{theorem}{Theorem}

\newtheorem{definition}{Definition}

\newcommand{\selma}[1]{{\color{black}#1}}

\begin{document}
\title{Fundamental Limits of Simultaneous Energy and Information Transmission
	}
	
%
%
\author{
\IEEEauthorblockN{Selma Belhadj Amor}
		\IEEEauthorblockA{INRIA, Lyon, France\\
			Email: selma.belhadj-amor@inria.fr}
		\and \IEEEauthorblockN{Samir M. Perlaza}
		\IEEEauthorblockA{INRIA, Lyon, France\\
			Email: samir.perlaza@inria.fr}

\thanks{This research is supported in part by the European Commission under Marie Sk\l{}odowska-Curie Individual Fellowship No.~659316 (CYBERNETS).}	
%
}

\maketitle

\begin{abstract}	
In this paper,  existing results regarding the fundamental limits of simultaneous energy and information transmission in wireless networks are reviewed. 
In point-to-point channels, the fundamental limits on the information rate given a minimum energy rate constraint are fully characterized by the notion of \emph{information-energy capacity function} introduced by Varshney in 2008. In a centralized multi-user channel, the fundamental limits on the information rates given a minimum energy rate constraint are described by the notion of \emph{information-energy capacity region} \selma{introduced by Fouladgar and Simeone in 2012}. Alternatively, in a decentralized multi-user channel, \selma{the authors have recently introduced the notion of  \emph{information-energy Nash region} to describe  these fundamental limits}. All these fundamental limits reveal the intrinsic trade-off between the \selma{two} conflicting tasks of information and energy transmission.
\end{abstract}

\section{Introduction}
Efficient energy utilization is among the main challenges of future communication networks in order to extend their lifetime and to reduce operating costs. Networks rely generally on battery-dependent devices. In some cases, such battery-dependency is relevant \selma{to such an extent that} the battery lifetime is the network lifetime as well. This is typically the case of wireless sensor networks. Once sensors are deployed, their batteries become generally inaccessible and cannot be recharged or replaced. Within this context, wireless energy transmission becomes an alternative to eliminate the need for \emph{in situ} battery recharging.
Nonetheless, for decades, the traditional engineering perspective was to design separately information transmission systems and energy transmission systems.
However, this approach has been shown to be suboptimal~\cite{VAR12} due to the fact that a radio frequency (RF) signal carries both energy and information. 
From this standpoint, a variety of modern wireless systems and proposals question the conventional separation approach and suggest that RF signals can be  simultaneously used for information and energy transmission~\cite{Bi15}. 

Typical examples of communications technologies already exploiting this principle 
are radio frequency identification (RFID) devices and 
power line communications. 
Beyond the existing applications, simultaneous energy and information transmission (SEIT) appears as a promising technology for a variety of emerging applications including low-power short-range communication systems, sensor networks, machine-to-machine networks and body-area networks, among others~\cite{KRI1}. 

When a communication system involves sending energy along with information, it should be designed to simultaneously meet two goals:
$(a)$ To reliably transmit energy at a given rate with a sufficiently small probability of energy outage; and
$(b)$ To reliably transmit information at a given rate with a sufficiently small probability of error.
However, these two tasks are usually conflicting. In fact, from a global perspective,  imposing a constraint on an energy rate impacts deeply the overall performance of the network components. 
%
%
An object of central interest in this perspective is the \emph{information-energy capacity region} which is the set of all information-energy rate tuples at which reliable transmission of information/energy is possible. \selma{The achievability of such tuples is subject to the existence of central controller that determines a network operating point and imposes the transmit or receive configuration that should be adopted by each network component. 
	However, in a decentralized network, each network component is an autonomous decision maker that aims to maximize its own individual reward by appropriately choosing a particular transmit or receive configuration. From this perspective, the individual choice of each component  does not necessarily achieve the capacity of the network. Hence, only the information-energy rate tuples that are \emph{stable} can be possible asymptotic operating points in a decentralized network. Stability can be understood in the sense of Nash~\cite{Nash-PNAS-1950}. That is, none of the network components is able to increase its own reward by unilaterally changing its own configuration.} 
This paper reviews the existing results regarding the fundamental limits of SEIT in wireless networks. These fundamental limits are characterized in terms of \emph{information-energy capacity function} \selma{introduced by Varshney \cite{VAR}} in point-to-point channels  and in terms of \emph{information-energy capacity region} \selma{introduced by Fouladgar and Simeone \cite{Fouladgar-CL-2012}} in centralized multi-user channels. Alternatively, in a decentralized multi-user channel, these fundamental limits are described by the \emph{information-energy Nash region} \selma{recently introduced by the authors \cite{Belhadj-amor-CISS-2016}}. \selma{The main focus will be on the Gaussian multi-access channel as a basic setup to capture the key aspects of the intrinsic information-energy trade-off in multi-user channels. However, most of the results continue to hold for more general network structures.}

\section{Point-to-point Information-Energy Trade-off}
\label{sec:p2p}
 In a point-to-point communication, information and energy transmission are subject to a trade-off between the information rate (bits per channel use) and the energy rate (energy units per channel use). \selma{This }is evidenced for instance in the constraints induced by the choice of a given modulation~\cite{Popovski}. Consider the transmission of a $4$-PAM signal over a point-to-point channel in the alphabet $\lbrace -2, -1, 1, 2\rbrace$. If there is no received energy rate constraint, one can clearly convey up to $2$ bits/ch.use by using all symbols of the constellation. However, if one requires the received energy rate to be for instance the maximum possible, the maximum transferable information rate is 1 bit/ch.use using only the most energetic symbols.
 \subsection{Discrete Memoryless Channels}
%
A \emph{discrete memoryless channel (DMC) with energy harvester (EH)} 
is characterized by a finite input alphabet $\set{X}$, two finite output alphabets $\set{Y}$ and $\set{S}$, and a conditional probability distribution $P_{YS|X}$. 
Let $n$ denote the transmission blocklength. 
At each time $t\in \{1,\dots,n\}$, if the input symbol $x_t\in \set{X}$ is transmitted, the probability of observing the channel output $y_t\in \set{Y}$ at the receiver and the additional output $s_t\in \set{S}$ at the EH is $P_{YS|X}(y_t,s_t|x_t)$. 
In the following, there is no particular assumption on \selma{the} joint distribution $P_{YS}$.

Within this context, two main tasks are to be simultaneously accomplished: information transmission and energy transmission.

\subsubsection{Information Transmission}
The goal of information transmission is that, over $n$ \selma{channel uses}, the transmitter conveys a message $M$ to the receiver at rate $R$ bits per channel use. 
The message $M$ is uniformly distributed over the set $\set{M}\eqdef \{1,\dots, \lfloor 2^{nR}\rfloor\}$. The channel input at time $t$ is
\begin{IEEEeqnarray}{rCl}
	\label{eq:enc}
	X_t=\varphi_t^{(n)}(M),\quad t\in \{1,\dots,n\},
\end{IEEEeqnarray}
for some \emph{encoding function} $\varphi_t^{(n)}$ of the form $\varphi_t^{(n)}\colon  \set{M} \to \mathbb{R}$.

The receiver observes the random sequence $Y^n$ and uses it to estimate the message $M$ by means of an appropriate decoding rule $\hat{M}^{(n)}=\Phi^{(n)}(Y^n)$ of the form $\Phi^{(n)}\colon \set{Y}^n\to \set{M}$\selma{. The} average probability of error is given by
\begin{equation}
P_{\error}^{(n)} (R) \eqdef \Pr(\hat{M}^{(n)}\neq M).
\end{equation}

\subsubsection{Energy Transmission}
At each time $t$, the energy that can be harvested from the output letter $s_t$ is given by $\omega(s_t)$ for \selma{a given} energy function $\omega: \set{S}\to\Reals_+$. For the $n$-length sequence $s^n$, the \selma{ harvested energy} is
$\omega(\vect{s}^n)=\sum_{t=1}^n \omega(s_t)$
(Note that the \selma{harvested energy can also be written as a function} of the input sequence).
The expected energy rate (in energy-units per channel use) at the EH is given by
\begin{equation}
B^{(n)}\eqdef \frac1n\sum_{t=1}^n \selma{\E[S]{\omega(S_t)}}.
\end{equation}
The goal of \selma{energy} transmission is to guarantee that the expected energy rate $B^{(n)}$ is not less than a given target energy transmission rate $B$ \selma{satisfying} 
$0 < B \leqslant \selma{\underline{B}}$. Here   $\selma{\underline{B}}$ is the maximum feasible energy rate.
\selma{The} probability of energy outage is defined as follows:
\begin{IEEEeqnarray}{rCl}
	\label{EqEnergyOutage}
	P_\outage^{(n)} (B) &=&\Pr\big\{B^{(n)}<B-\epsilon\big\},
\end{IEEEeqnarray}
for some $\epsilon>0$ arbitrarily small.

\subsubsection{Simultaneous Energy and Information Transmission}
%
%
%
%

The DMC 
is said to operate at the information-energy rate pair $(R,B) \in \mathds{R}_{+}^{2}$ when the  transmitter and the receiver use a transmit-receive configuration such that:
$(i)$ information transmission occurs at rate $R$ with probability of error arbitrarily close to zero; and 
$(ii)$ energy transmission occurs at a rate not smaller than $B$ with energy-outage probability arbitrarily close zero.
Under these conditions, the information-energy rate pair $(R,B)$ is said to be \emph{achievable}. 
\begin{definition}[Achievable Information-Energy Rates]\label{DefAchievablePairs}
The  $(R,B) \in \mathds{R}_{+}^{2}$ is achievable if there exists a sequence of encoding and decoding functions  $\big\{\{\selma{\varphi}_{t}^{(n)}\}_{t=1}^n,\Phi^{(n)}\big\}_{n=1}^\infty$ such that both the average error probability and the energy-outage probability tend to zero as the blocklength $n$ tends to infinity. That is,
\begin{IEEEeqnarray}{lcl}
\label{EqProbError}
\limsup_{n \rightarrow \infty}\;  P_{\error}^{(n)}(R)  & = & 0, \mbox{ and }\\
\label{EqProbPower}
\limsup_{n \rightarrow \infty}\;  P_\outage^{(n)} (B)& = & 0.
\end{IEEEeqnarray}
\end{definition}
Often, increasing the energy rate implies decreasing the information \selma{rate} and \emph{vice versa}.
This trade-off is accurately modeled by the notion of \emph{information-energy capacity function}. The goal is set to maximize \selma{the information rate under a minimum received energy rate constraint.}
\begin{definition}[Information-Energy Capacity Function]\label{DefCEF}
	Let $b\in [0,\selma{\underline{B}}]$ denote the minimum energy rate that must be guaranteed at the input of the \selma{EH}. For each blocklength $n$, define the function $C^{(n)}(b)$ as follows:
	\begin{equation}
	C^{(n)}(b)\eqdef \max_{X^n: B^{(n)}\geqslant b} I(X^n;Y^n),
	\end{equation}
	\selma{where $I(X^n;Y^n)$ denotes the mutual information between $X^n$ and $Y^n$ and }the maximization is over all the length-$n$ input sequences $X^n$ for which the expected energy rate $B^{(n)}$ is not smaller than $b$. The information-energy capacity function for a minimum energy rate $b$ is defined as
		\begin{equation}
		C(b)\eqdef \limsup_{n\to \infty} \quad \frac1n C^{(n)}(b).
		\end{equation}
\end{definition}
\begin{theorem}[Information Capacity Under Minimum Energy Rate (Theorem~1 in \cite{VAR})]
The supremum over all achievable information rates in the DMC under a minimum energy rate $b$ in energy-units per channel use is given by $C(b)$ in bits per channel use.  
\end{theorem}
\subsubsection{Examples}
This subsection reviews some \selma{closed-form} expressions (provided in \cite{VAR}) of \selma{the} information-energy capacity function \selma{to better  illustrate the optimal trade-offs between information and energy rates in the considered point-to-point channels.}
Three binary channels are considered for the special case in which the receiver and the EH observe the same output sequence, i.e., at each time $t$, $S_t=Y_t$. Thus, the channel law reduces to $P_{Y|X}$. 

In a noiseless binary channel, the information-energy capacity function for a minimum energy rate $b$ is 
\begin{equation}
C(b) = \left\lbrace
\begin{array}{lcl}
1, 	& \mbox{if }  	& 0\leqslant b\leqslant \frac12,\\
H_2(b), 			& \mbox{if }& \frac12\leqslant b\leqslant 1,
\end{array}\right.
\end{equation}
where $H_2(\cdot)$ is the binary entropy function. 
For any $0\leqslant b \leqslant \frac12$ the energy rate constraint is vacuous and equiprobable inputs achieve capacity. However, when $\frac12\leqslant  b \leqslant 1$, the capacity-achieving distribution is Bernouilli with parameter $b$. Note that the capacity is monotonically decreasing with $b$\selma{. Hence,} the more energy is requested, the more the transmitter is forced to use the most energetic symbol which reduces the information rate.

In a binary symmetric channel with cross-over probability~$p$, the information-energy capacity function for a minimum energy rate $b$ is 
\begin{equation}
C(b) = \left\lbrace
\begin{array}{lcl}
1-H_2(p), 	& \mbox{if }  	& 0\leqslant b\leqslant \frac12 \text{ and }\\
H_2(b)-H_2(p), 			& \mbox{if }& \frac12\leqslant b\leqslant 1-p.
\end{array}\right.
\end{equation}
In the energy-unconstrained problem, equiprobable inputs are
capacity-achieving and yield an energy rate of $\frac12$. For any $0\leqslant b \leqslant \frac12$\selma{,} the energy rate constraint is vacuous. For $b>\frac12$, the distribution must be
perturbed so that the symbol $1$ is transmitted more frequently to increase the energy rate. The maximum energy rate that is feasible
is $1-p$, when $1$ is always transmitted.

In the Z-channel with $1$ to $0$ cross-over probability $\epsilon$, i.e., the binary DMC with $P_{Y|X}=\begin{bmatrix}
1&0\\
\epsilon& 1-\epsilon
\end{bmatrix}$, the information-energy capacity function for a minimum energy rate $b$ is 
\begin{equation}
C(b) = \left\lbrace
\begin{array}{lcl}
C(0),
& \mbox{if }  	& 0\leqslant b\leqslant (1-\epsilon)\pi^*  \text{ and }\\
H_2(b)-\frac{b}{1-\epsilon} H_2(\epsilon), 			& \mbox{if }& (1-\epsilon)\pi^* \leqslant b\leqslant 1-\epsilon,
\end{array}\right.
\end{equation}
\selma{where $C(0)$ denotes the capacity of this channel~\cite{Zchannel} given by
\begin{equation}
C(0)=\log_2\left(1-\epsilon^{\frac{1}{1-\epsilon}}+\epsilon^{\frac{\epsilon}{1-\epsilon}}\right).
\end{equation}
The capacity-achieving input distribution is Bernouilli with parameter
\begin{equation}
\pi^*=\frac{\epsilon^{\frac{\epsilon}{1-\epsilon}}}{1+(1-\epsilon) \epsilon^{\frac{\epsilon}{1-\epsilon}}}.
\end{equation}}

The three examples show that the more stringent the energy rate constraint is, the more the transmitter needs to adapt its optimal strategy and switch over to using the most energetic symbol.

%

\subsection{Gaussian Memoryless Channel}
\label{sec:Gauss}
The results in \cite{VAR} extend directly to \selma{memoryless continuous} alphabet channels. 
In the memoryless Gaussian channel with EH, 
 at each channel use $t\in \{1,\dots,n\}$, if $X_{t}$ denotes the real \selma{symbol} sent by the transmitter, the receiver observes the real channel output 
\begin{equation}
\label{EqY}
Y_{t}=h_{1} X_{t} + Z_t,
\end{equation}
and the EH observes 
\begin{equation}
\label{EqS}
S_{t} = h_{2} X_{t} + Q_t,
\end{equation}
with $h_{1}$ and $h_{2}$ constant non-negative channel coefficients satisfying the $\set{L}_2$-norm condition:
$\|\vect{h}\|^2 \leqslant 1$, 
with $\vect{h}\eqdef \trans{(h_{1},h_{2})}$ to ensure the principle of \selma{energy conservation}.
The noise sequences $Z_t$ and $Q_t$ are identically distributed standard real Gaussian variables. 
The output energy function for this channel is given by $\omega(s)\eqdef s^2$ and the input sequence \selma{$\{X_t\}_{t=1}^n$} satisfies an average \emph{input power constraint}
\begin{IEEEeqnarray}{C}
	\label{EqPowerConstraint}
	\frac1n \sum_{t=1}^n \E{X_{t}^2} \leqslant P,
\end{IEEEeqnarray}
with $P$ the average \selma{transmit power}  in energy-units per channel use.
The channel is fully described by the signal to noise ratios (SNRs) defined as $\SNR_i\eqdef h_i^2 P$, for $i \in \{1,2\},$ given the normalization over the noise powers. 

In the memoryless Gaussian channel\selma{,} the alphabets are \emph{continuous}. Nonetheless, \selma{information transmission and energy transmission} can be described similarly to the DMC (where the finite input and output alphabets are replaced by $\Reals$). The achievability, the information-energy capacity region, and the information capacity function can be defined similarly to the DMC when taking into account the average input power constraint $P$.  

The information capacity of the Gaussian channel without minimum energy rate constraint is $\set{C}(0,P)=\frac12 \log_2(1+\SNR_1)$  and the maximum energy rate which can be achieved at the input of the EH is $\selma{\underline{B}}\eqdef1 + \SNR_{2}$.

For any $0\leqslant b \leqslant 1+\SNR_2$, the information-energy capacity function is
\begin{equation}
\set{C}(b,P)=\max_{
	X:\E{X^2}\leq P \text{ and } \E{S^2}\geqslant b  }\quad I(X;Y).
\end{equation} 

Following similar steps as in \cite{VAR} and \cite{SMITH}, it can be shown that 
\begin{equation}
\set{C}(b,P)=\frac12 \selma{\log_2}\left(1+\SNR_1\right),
\end{equation}
which equals the capacity of the Gaussian channel without EH under average input-power constraint $P$, achieved using zero-mean Gaussian inputs with variance $P$. Thus in this case for any feasible energy rate $0\leqslant b \leqslant 1+\SNR_2$ the information-optimal strategy is unchanged.

In the Gaussian channel with peak power constraint, depending on the value of the amplitude constraint, a trade-off between the information and energy rates may be observed~(See~\cite{VAR} and \cite{SMITH}).

\section{Multi-User Simultaneous Energy and Information Transmission}
Unlike point-to-point setups, multi-user SEIT requires generally additional transmitter cooperation/coordination to increase the energy rate at the input of the EH. \selma{Consider a network in which a given transmitter simultaneously transmits energy to an EH and information to a receiver.} If this transmitter is required to deliver an energy rate that is less than what it is able to deliver by only transmitting information, it is able to fulfill the task independently of the behavior of the other transmitters since  it can use all its power budget to maximize its information transmission rate and it \selma{is still} able to meet the energy rate constraint. 
Alternatively, when the requested energy rate is higher than what it is able to deliver by only transmitting information, its behavior is totally dependent on the behavior of the other transmitters. 
 In this case, the minimum energy rate constraint drastically affects the way that the transmitters interact with each other. More critical scenarios are the case in which the requested energy rate is \selma{more} than what all transmitters are able to deliver by simultaneously transmitting information using all the available individual power budgets.  In these cases, none of them can unilaterally ensure reliable energy transmission at the requested rate. Hence, \selma{the} transmitters must engage in a mechanism through which an energy rate that is higher than the energy delivered by exclusively transmitting information-carrying signals is ensured at the EH. This suggests for instance, the use of power splits in which the transmitted symbols have an information-carrying and an energy-carrying components. The latter typically consists \selma{of} signals that are known at all devices and can be constructed such that the energy captured at the EH is maximized. 
Moreover, the information-energy trade-off takes different facets depending on whether or not the network is centralized. In the former, there exists a central controller that determines an operating point and indicates to each transmitter and its corresponding receiver(s) the appropriate transmit-receive configuration to achieve such a point. In the latter, each network component is considered to be autonomous and seeks to determine its own transmit-receive configuration in order
to maximize its individual benefit.
Clearly, the operating points of the network are significantly different depending on the degree of control over all devices. That is, in a centralized network, all achievable information-energy rates are feasible operating points as the base-station can impose a particular operating point via a signaling system. However, in a fully decentralized network, only stable operating points are feasible, as each device tunes its transmit-receive configuration aiming \selma{at} maximizing its own individual performance. 

To understand the optimal behavior of SEIT in a multi-user network, an important scenario to look at is the multi-access channel (MAC) with an EH. From an information theoretic viewpoint, the information-energy trade-off was studied by Fouladgar \textit{et al.}~\cite{Fouladgar-CL-2012} in the discrete memoryless two-user MAC. \selma{Recently, Belhadj Amor \textit{et al.} studied  SEIT in the centralized Gaussian MAC (G-MAC) with and without channel-output feedback~\cite{InriaRep, BelhadjAmor-COMNET-2015} as well as in the decentralized G-MAC~\cite{Belhadj-amor-CISS-2016}.}

\subsection{Gaussian Multi-Access Channel}
\selma{Consider the channel model described in section~\ref{sec:Gauss} with two transmitters.}
 Transmitters 1 and 2 wish to send two independent messages $M_1$ and $M_2$ to the single receiver at rates $R_1$ and $R_2$. This channel model is called two-user memoryless Gaussian multiple access channel (G-MAC) with an EH. Transmitter $i$ has an input  power constraint $P_i$ and channel coefficients $h_{1i}$ and $h_{2i}$ to the receiver and the EH, respectively.
These channel coefficients satisfy the $\set{L}_2$-norm condition:
$\forall j\in \{1,2\},\quad \|\vect{h}_j\|^2 \leqslant 1$,
with $\vect{h}_j\eqdef \trans{(h_{j1},h_{j2})}$ in order to meet the energy conservation principle and $\SNR_{ji}$, with $\forall (i,j) \in \{1,2\}^2$ are defined as:
$\SNR_{ji}\eqdef |h_{ji}|^2 P_{i}.$
%
Encoding, decoding, probability of error, probability of energy outage, and  achievability can be defined analogously to the Gaussian point-to-point channel when taking into account the considerations described above. 
\selma{
The energy rate at the input of the EH cannot exceed the maximum feasible value given by $\selma{\underline{B}}\eqdef1 + \SNR_{21} +\SNR_{22} + 2\sqrt{\SNR_{21} \SNR_{22}}$. This can be achieved when the transmitters use all their power budgets to send fully correlated channel inputs.
	}
In the sequel, 	let $b \in [0,1 + \SNR_{21} +\SNR_{22} + 2\sqrt{\SNR_{21} \SNR_{22}}]$ be the minimum energy rate required at the input of the EH.

\subsection{Centralized SEIT in G-MAC}
In a centralized G-MAC, the fundamental limits of the information-energy trade-off are fully characterized by the information-energy capacity region with a minimum energy rate constraint $b$, i.e., the closure of all achievable information-energy rate triplets $(R_1,R_2,B)$, is described by the following theorem.
\begin{theorem}[Information-Energy Capacity Region with Minimum Energy Rate $b$ (Theorem~1 in~\cite{Belhadj-amor-CISS-2016})] 
The information-energy capacity region $\set{E}_{b}(\SNR_{11},\SNR_{12},\SNR_{21},\SNR_{22})$ of the G-MAC with minimum energy rate constraint $b$ is given by the set of all non-negative information-energy rate triplets $(R_1, R_2, B)$ that satisfy
	\begin{subequations}
		\label{eq:regprop1}	
		\begin{IEEEeqnarray}{llclll}
			& &R_1 & \leqslant & \frac{1}{2} \log_2\left( 1 + \beta_1 \SNR_{11}  \right)\label{eq:thm1_c1},\\
			& & R_2 & \leqslant & \frac{1}{2} \log_2\left( 1 + \beta_2 \SNR_{12} \right)\label{eq:thm1_c2},\\
			& & R_1 + R_2 & \leqslant & \frac{1}{2} \log_2 \big( 1 +\beta_1\SNR_{11} + \beta_2\SNR_{12}\big), \IEEEeqnarraynumspace\label{eq:thm1_c12}\\
			b&\leqslant & B  & \leqslant &  1 + \SNR_{21} +  \SNR_{22}\nonumber\\
			& &&&+ 2  \sqrt{(1-\beta_1)\SNR_{21} (1-\beta_2)\SNR_{22}},\label{eq:thm1_e}
		\end{IEEEeqnarray}
	\end{subequations}
	with $(\beta_1,\beta_2) \in \left[ 0, 1 \right]^2 $. 
\end{theorem}
The terms $\beta_1$ and $\beta_2$ in \eqref{eq:regprop1} might be interpreted as the fractions of power that transmitter $1$ and transmitter $2$ allocate for information transmission, respectively. The remaining fraction of power $(1-\beta_i)$ is allocated by transmitter $i$ for  exclusively transmitting energy to the EH by sending common randomness known non-causally to all terminals. 
For any $(R_1,R_2,B)$, whenever the  energy rate $B$ is smaller than the energy rate required to guarantee reliable communications at the information rates $R_1$ and $R_2$, the
energy rate constraint is vacuous since it is always satisfied and each transmitter can exclusively use its available power budget
to increase its information rate, i.e, $\beta_1=\beta_2=1$. Alternatively, when the energy rate $B$ is higher than what is strictly necessary to guarantee reliable communication, the transmitters face a trade-off between information and energy rates. Often, increasing the energy rate implies decreasing
the information rates and \emph{vice-versa}.
%
\subsection{Decentralized SEIT in G-MAC}


In a decentralized G-MAC,  the aim of transmitter $i$, for $i\in \{1,2\}$, is to autonomously choose its transmit configuration $s_i$ in order to maximize its information rate $R_i$, while guaranteeing a minimum energy rate $b$ at the EH.
The receiver is assumed to adopt a fixed decoding strategy that is known in advance \selma{to} both transmitters.  
%
The choice of the transmit configuration of  each transmitter is subject to the choice of the other transmitter as both of them must guarantee the minimum energy constraint; and at the same time, depending on the decoding scheme at the receiver, the information-carrying signal of one transmitter is interference to the other transmitter.

The competitive interaction of the two transmitters and the receiver in the decentralized G-MAC with minimum energy constraint $b$  can be modeled by the following game in normal form:
$\GameNF$,
where $b$ is a parameter of the game, 
 the set $\mathcal{K} = \lbrace 1, 2 \rbrace$ is the set of players (transmitters $1$ and  $2$), and the sets $\mathcal{A}_1$ and $\mathcal{A}_2$ are their sets of actions. An action  $s_i \in \mathcal{A}_i$ of a player $i \in \mathcal{K}$ is basically its transmit  configuration. 
The utility function of transmitter~$i$ is $u_i: \mathcal{A}_1 \times \mathcal{A}_2 \rightarrow \mathds{R}_+$ and it is defined as
\begin{equation}
\label{EqUtilityTX}
\small
u_i(s_1,s_2) = \left\lbrace
\begin{array}{lcl}
\hspace*{-2mm}R_i( s_1,s_2), &\hspace*{-4mm} \mbox{if }\hspace*{-3mm}  &\hspace*{-2mm} P_{\error}^{(n)}(\selma{R_1,R_2}) < \epsilon \text{ and } P_\outage^{(n)}(\selma{b})< \delta\\
\hspace*{-2mm}-1, & &\hspace*{-2mm}\mbox{otherwise,}
\end{array}\right.
\end{equation}
where $\epsilon > 0$ and $\delta > 0$ are arbitrarily small numbers and $R_i(s_1,s_2)$ (written as $R_i$ for simplicity) denotes an information rate achievable  with the configurations $s_1$ and $s_2$.
\selma{
	Note that the utility is -1 when either the error probability or the energy outage probability is not arbitrarily small. This
	is meant to favor the action profiles in which there is no information transmission (information rate and error probability are zero) but there is energy transmission (probability of energy outage can be made arbitrarily close to zero), over the actions in which the information rate is zero but the energy constraint is not satisfied.
	}
Note  that there might exist several transmit configurations that achieve the same triplet $(R_1, R_2, B)$ and distinction is made only when needed. 

The fundamental limits of SEIT in the decentralized G-MAC are fully characterized the $\eta$-Nash equilibrium \cite{Nash-PNAS-1950} ($\eta$-NE) information-energy region, with $\eta \geqslant 0$ arbitrarily small. This region corresponds to the set of information-energy rate triplets $(R_1,R_2,B)$ that are achievable and \emph{stable} in the G-MAC where stability is considered in the sense of Nash~\cite{Nash-PNAS-1950}. More specifically, an action profile (a transmit configuration)  $(s_1^*,s_2^*)$ is an $\eta$-NE, if none of the transmitters can increase its own information rate  by  more than $\eta$ bits per channel use by changing its own transmit configuration and keeping the average error probability and the energy outage probability arbitrarily close to zero.

%
%
The $\eta$-NE information-energy region of the game $\gameNF$ when the receiver uses single-user decoding (SUD), denoted by $\set{N}_{\mathrm{SUD}}(b)$, is described by the following theorem. 
\begin{theorem}[$\eta$-NE Information-Energy Region of the Game $\gameNF$ with SUD (Theorem~2 in \cite{Belhadj-amor-CISS-2016})]\label{thm1}
%
%
The set  $\set{N}_{\mathrm{SUD}}(b)$ of $\eta$-NEs of the game $\gameNF$ is contains all  information-energy rate-triplets $(R_1,R_2,B)$ which satisfy:
	\begin{subequations}	\label{cstThm1}
\begin{IEEEeqnarray}{lclcl}
			0&\leqslant & R_1 &=& \frac12 \log_2\left(1+\frac{\beta_1\SNR_{11}}{1+\beta_2\SNR_{12}}\right),\\
			0 &\leqslant & R_2 &=&  \frac12 \log_2\left(1+\frac{\beta_2\SNR_{12}}{1+\beta_1\SNR_{11}}\right),\\
			b&\leqslant	&B     & \leqslant & 1 + \SNR_{21} +  \SNR_{22}\nonumber\\&&&&+ 2  \sqrt{(1-\beta_1)\SNR_{21} (1-\beta_2)\SNR_{22}},
		\end{IEEEeqnarray}
	\end{subequations}
	where $\beta_1=\beta_2=1$ when $b\in [0,1+\SNR_{21}+\SNR_{22}]$ and
	$(\beta_1,\beta_2)$ satisfy
	\begin{equation}
	b=1 + \SNR_{21} +  \SNR_{22}
	+ 2  \sqrt{(1-\beta_1)\SNR_{21} (1-\beta_2)\SNR_{22}}
	\end{equation}
	when $b\in (1+\SNR_{21}+\SNR_{22},1+\SNR_{21}+\SNR_{22}+2 \sqrt{\SNR_{21}\SNR_{22}} ]$. 
\end{theorem}

Let ${\mathrm{SIC}(i\rightarrow j)}$ denote the case in which the receiver uses successive interference cancellation (SIC) with decoding order: transmitter $i$ before transmitter $j$, with $i \in \lbrace 1,2 \rbrace$. In this case, the $\eta$-NE information-energy region of the game $\gameNF$, denoted by $\set{N}_{\mathrm{SIC}(i\rightarrow j)}(b)$,  is described by the following theorem.
\begin{theorem}[$\eta$-NE Information-Energy Region of the Game $\gameNF$ with SIC (Theorem~3 in \cite{Belhadj-amor-CISS-2016})]\label{thm2}
	The set $\set{N}_{\mathrm{SIC}(i\rightarrow j)}(b)$ contains all information-energy rate-triplets $(R_1,R_2,B)$ satisfying:
	\begin{subequations}	\label{cstThm2}
		\begin{IEEEeqnarray}{lclcl}
			0&\leqslant & R_i &=& \frac12 \log_2\left(1+\frac{\beta_i\SNR_{1i}}{1+\beta_j\SNR_{1j}}\right), \\
			0&\leqslant & R_j &=& \frac12 \log_2\left(1+ \beta_j\SNR_{1j} \right),\\
			b &\leqslant & B   &\leqslant &   1 + \SNR_{21} +  \SNR_{22}\nonumber\\&&&&+ 2  \sqrt{(1-\beta_1)\SNR_{21} (1-\beta_2)\SNR_{22}},
		\end{IEEEeqnarray}
	\end{subequations}
	where $\beta_1=\beta_2=1$ when $b\in [0,1+\SNR_{21}+\SNR_{22}]$ and
	$(\beta_1,\beta_2)$  satisfy
	\begin{equation}
	b=1 + \SNR_{21} +  \SNR_{22}
	+ 2  \sqrt{(1-\beta_1)\SNR_{21} (1-\beta_2)\SNR_{22}}
	\end{equation}
	when $b\in (1+\SNR_{21}+\SNR_{22},1+\SNR_{21}+\SNR_{22}+2 \sqrt{\SNR_{21}\SNR_{22}} ]$.
%
%
\end{theorem}
\selma{Let $\set{N}(b)$ denote the $\eta$-NE region of the game $\gameNF$ when the receiver uses any time-sharing  between the previous decoding techniques. The set $\set{N}(b)$ is defined as:
		\begin{equation}
		\set{N}(b)=\textnormal{Closure }\hspace*{-1.5mm}\biggr(\hspace*{-1mm}\set{N}_{\mathrm{SUD}}(b) \cup \set{N}_{\mathrm{SIC}(1\rightarrow 2)}(b) \cup \set{N}_{\mathrm{SIC}(2\rightarrow 1)}(b)\hspace*{-1mm}\biggr)\hspace*{-.5mm}.	
		\end{equation}
	That is, if the receiver performs any time-sharing combination between SUD, SIC$(1\to 2)$, and SIC$(2\to 1)$ then the transmitters can use the same time-sharing combination between their corresponding $\eta$-Nash equilibria strategies to  achieve any point inside $\set{N}(b)$.
	}

\begin{figure}[ht]
	\hspace*{-3mm}
	\centering{
			\includegraphics[scale=0.365]{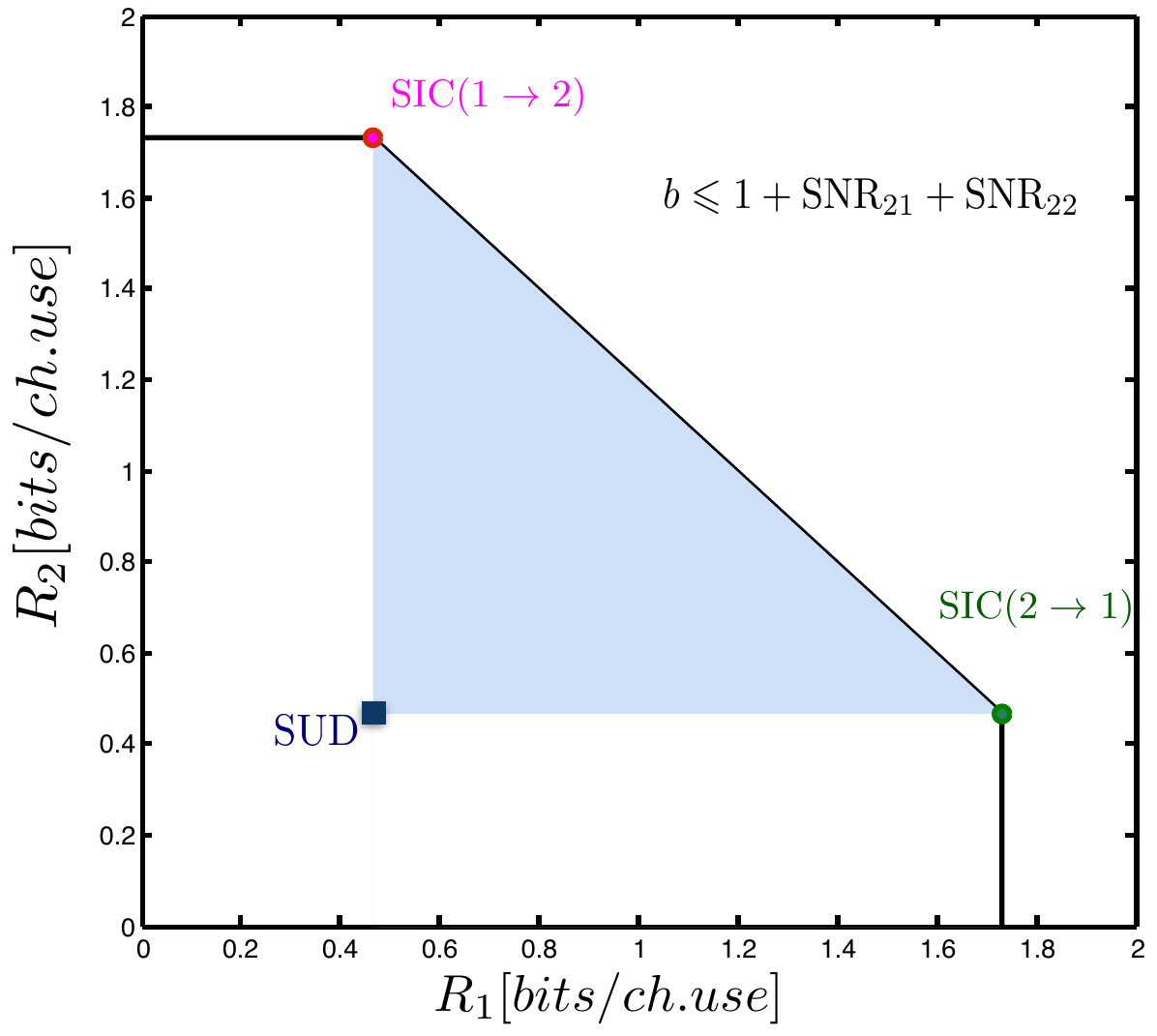}
			\includegraphics[scale=0.33]{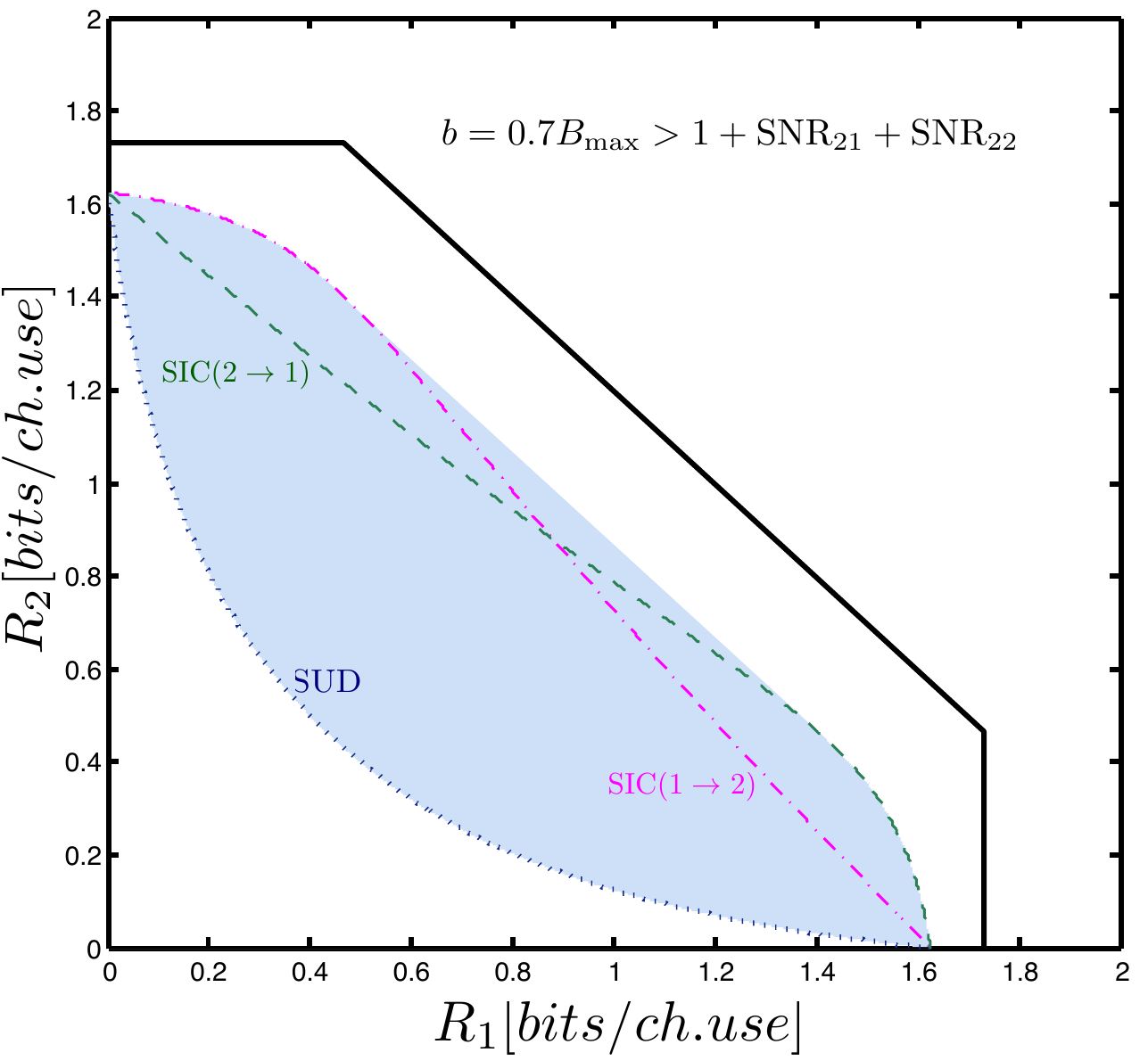}	\vspace*{-6mm}
			}	
		\caption{	\label{fig:dist_regs}
			Projection of the sets \selma{ $\set{N}_{\mathrm{SUD}}(b)$, $\set{N}_{\mathrm{SIC}(i\rightarrow j)}(b)$, and $\set{N}(b)$ (blue region)}  over the $R_1$-$R_2$ plane for different values of  $b$.
			The information capacity region is also plotted as a reference (white region inside solid lines) for $\SNR_{11}=\SNR_{12}=\SNR_{21}=\SNR_{22}=10$. }
		\vspace*{-6mm}
\end{figure}

Fig.~\ref{fig:dist_regs} shows the projection of the regions described in Theorem \ref{thm1}  and Theorem \ref{thm2} as well as the convex hull of these regions for a symmetric G-MAC with $\SNR_{11} = \SNR_{12} = \SNR_{21} = \SNR_{22} = 10$ (EH and receiver are co-located). Note that for all $b \leqslant 1 + \SNR_{21} + \SNR_{22}$,  both transmitters use the whole available power for information transmission (see the figure on the left).
Alternatively, when $b > 1 + \SNR_{21} + \SNR_{22}$, both transmitters use the minimum energy needed to make the energy-outage probability arbitrarily close to zero and seek for the largest possible information transmission rates (See the figure on the right). 
\selma{One of the main observations is that the existence of an $\eta$-NE, with $\eta$ arbitrarily small, is always guaranteed as long as the SEIT problem is feasible, that is, $b \leqslant 1+\SNR_{21}+\SNR_{22}+2 \sqrt{\SNR_{21}\SNR_{22}}$ since $\set{N}(b) \neq \emptyset$.}
%
\section{Discussion}
In point-to-point channels, depending on the channel model, the trade-off between information and energy rates is not always observed (\emph{e.g.}, Gaussian channel with peak power constraint~\cite{VAR,SMITH}).

In G-MACs, SEIT induces additional transmitter cooperation to meet the energy rate constraints. This cooperation is usually not natural especially when the transmitters do not share common information and are not co-located. In this sense, it seems likely that providing additional means of cooperation would result in a significant performance enhancement of SEIT. From this standpoint, exploring the benefits induced by cooperation techniques such as channel-output feedback and conferencing in SEIT for the two-user G-MAC is really promising, especially in terms of energy transmission. 
Recently, Belhadj Amor \textit{et al.} have shown that channel-output feedback can provide a multiplicative factor to the energy rate without any decrease on the information rates \cite{InriaRep}. 
\selma{This surprising result is mainly due to the additional correlation that can be induced among the signals of all the transmitters via feedback in order to increase the energy that can be collected at a given EH. This reasoning is valid for multi-access channels with an arbitrary number of users and it also applies to other multi-user setups such as the broadcast and interference channels.	}


\end{document}